\documentclass[10pt,journal,compsoc]{IEEEtran}
\usepackage{amssymb}
\usepackage{amsmath}
\usepackage{mathtools}
\usepackage{ellipsis}
\usepackage{textcomp}
\usepackage{cite}
\usepackage{graphicx}
\usepackage{epstopdf}
\usepackage{psfrag}
\usepackage{float}
\usepackage{grffile}
\usepackage{csquotes}
\usepackage{multicol}
\usepackage{subfigure}
\usepackage{epsfig}
\usepackage{multirow}
\usepackage{footnote}
\makesavenoteenv{tabular}
\makesavenoteenv{table}
\usepackage{pifont}
\usepackage{url}
\usepackage{color}

\usepackage{setspace}
\usepackage{amsthm}
\usepackage{array}

\usepackage{listings}
\usepackage{enumerate}
\usepackage[T1]{fontenc}
\usepackage{textcomp}
\usepackage{listings}
\UseRawInputEncoding

\usepackage{algorithm}
\usepackage[noend]{algpseudocode}
\usepackage{algorithmicx}
\algnewcommand\algorithmicoutput{\textbf{OUTPUT:}}
\algnewcommand\OUTPUT{\item[\algorithmicoutput]}

\algnewcommand\algorithmicinput{\textbf{INPUT:}}
\algnewcommand\INPUT{\item[\algorithmicinput]}

\algnewcommand\algorithmicinputs{\textbf{INPUTS:}}
\algnewcommand\INPUTS{\item[\algorithmicinputs]}

\algnewcommand\algorithmicproc{\textbf{PROCEDURE:}}
\algnewcommand\PROC{\item[\algorithmicproc]}

\algnewcommand\algorithmicending{\textbf{END}}
\algnewcommand\ENDING{\item[\algorithmicending]}

\algdef{SE}[DOWHILE]{Do}{DoWhile}{\algorithmicdo}[1]{\algorithmicwhile\ #1}

\begin{document}
\title{MSSI: Middleware for Unified Semantic\\ and Syntactic Interoperability in IoT}

\author{Sanku~Kumar~Roy,~\IEEEmembership{Student~Member,~IEEE}, 
	Sudip~Misra,~\IEEEmembership{Fellow,~IEEE},
	and~Narendra~Singh~Raghuwanshi
	\IEEEcompsocitemizethanks{\IEEEcompsocthanksitem S. K. Roy was affiliated with the Department of Computer Science and Engineering, Indian Institute of Technology, Kharagpur, 721302, India, during this work. He is currently with the Department of Computing Science at the University of Alberta, Edmonton, AB, Canada, T6G 2E8.\protect\\
	E-mail: sankukumarroy@gmail.com
	\IEEEcompsocthanksitem S. Misra is with the Department of Computer Science and Engineering, Indian Institute of Technology, Kharagpur, 721302, India.\protect\\
	E-mail: smisra@cse.iitkgp.ac.in
	\IEEEcompsocthanksitem  N S Raghuwanshi is with the Department of Agricultural and Food Engineering, Indian Institute of Technology, Kharagpur, 721302, India. \protect\\
	E-mail: nsr@agfe.iitkgp.ac.in}
	
}


\IEEEtitleabstractindextext{%
\begin{abstract}
	With the growing demand of Internet of Things (IoT), there is a need for seamless and reliable communication between heterogeneous IoT devices and the cyber-world to ensure autonomous control over any application process. More specifically, seamless communication requires interoperability between heterogeneous devices (actors) having different semantics and data formats (syntaxes), while making it more challenging. In this paper, we propose a middleware solution for unified semantic and syntactic interoperability in the publisher-subscriber framework of IoT network. The proposed framework automatically translates the subscribers (users) compatible syntax and semantics of the receiver message from the publishers (IoT devices). First, we propose a novel method of syntax translation of messages, to solve the syntactic disparities between users and devices, while providing the information in the user requested syntax. Thereafter, a multilayer perceptron (MLP)-based semantic interoperability framework is proposed to translate the device information to the user requested semantics. Additionally, a novel algorithm is proposed for extracting raw and discriminative features, which are to be fitted to the MLP model as inputs. To show the effectiveness of the proposed middleware, we evaluate different parameters, while considering various publicly used data formats and semantic annotations of attributes to ensure the versatility of the proposed middleware in the practical scenario. The overall classification accuracy using MLP is $95.78$\% for determining the standard meaning of each attribute of the incoming message from the publisher to address the semantic interoperability problem in IoT.
\end{abstract}
	
}

\maketitle
\IEEEdisplaynontitleabstractindextext
\IEEEpeerreviewmaketitle
\IEEEraisesectionheading{\section{Introduction}\label{prob3:sec:introduction}}
\IEEEPARstart{I}{nternet of Things} (IoT) is a vast network of interconnected sensors, devices, and systems bridging the digital and physical realms \cite{manogaran2022, mishra2023, Kaed2018}. IoT devices are manufactured by different vendors. These vendors use different syntaxes, structures, and semantics to represent device data both inside and outside any given context \cite{Kaed2018}. Thus, there is a requirement of seamless interoperability between all actors in the IoT network. As an example, if two devices want to communicate with each another, they should have a shared knowledge in terms of syntax and semantics of exchanged data. Achieving shared knowledge is a nontrivial task due to the existence of numerous technologies, protocols, and manufacturers at the communication of device.
\begin{figure}[t]
	\centering
	\includegraphics[width=8cm]{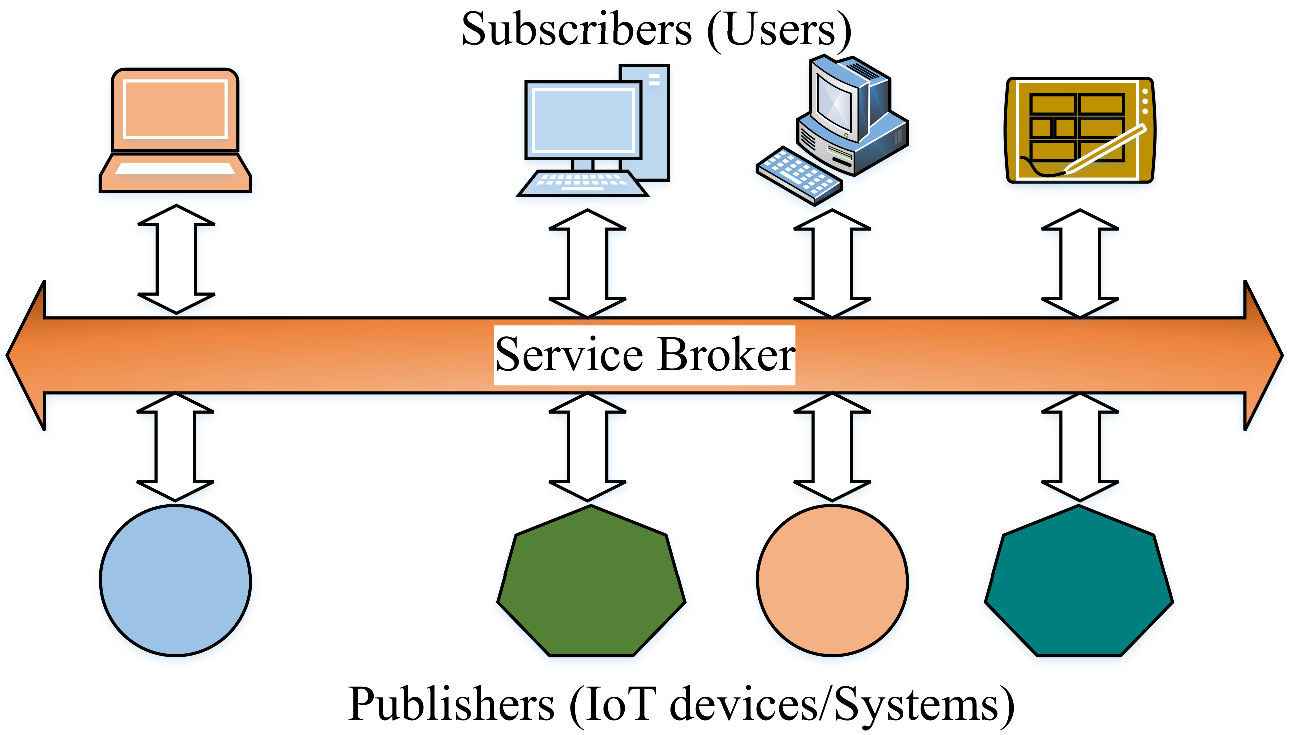}
	\caption{Publisher/Subscriber Framework}
	\label{prob3:fig:publish_subscriber_framework}
\end{figure}

In IoT, the popular communication protocols such as Message Queuing Telemetry Transport (MQTT), Constrained Application Protocol (CoAP), and Hypertext Transfer Protocol (HTTP) follow the publisher-subscribe framework (PSF), which is shown in Fig.\ref{prob3:fig:publish_subscriber_framework}. In this framework, both the publishers (IoT devices) and the subscribers (users) register themselves to the message data broker (server). The message data broker decouples the data transmission between the publishers and subscribers according to the service demand. 

In general, PSF uses standard data formats such as JavaScript Object Notation (JSON), Extensible Markup Language (XML), and Comma-Separated Values (CSV) as well as different semantic annotations. Therefore, to achieve seamless communication between a pair of device and user, there is a need of a translator between them, as different devices and users have different data format structures, i.e., different syntaxes, and use different attribute/element names to represent the same or different concepts, i.e. different semantics. The different data formats (syntaxes) and semantic annotations represent syntactic and semantic interoperability, respectively. To solve the problem mentioned above, the translator first interprets the shared information and then translates the sender information to the receiver compatible information format and semantics. So, it is clear that each pair of devices and users requires a specific translator for seamless communication between them. In the case of a large number of devices/users, the possible number of translators is even larger than the number of devices/users. For example, if $\mathcal{N}$ number of devices/users want to interact among themselves while each device uses a different semantic and format, the maximum number of required translators is $\mathcal{N}^2-\mathcal{N}$, which is even larger than the number of devices/users \cite{Moutinho2018}. This problem can be addressed if we introduce an automated translator, which converts the received messages to the compatible syntax and semantics of the users.

To address the problem, researchers\cite{Kopke2017,Nambi2014} focused on semantic interoperability. More specific, Xiao \textit{et al.}\cite{Xiao2014} focused on both semantic and syntactic interoperability between users and devices. K\"{o}pke \cite{Kopke2017} proposed the semantic translation method and semantic matching algorithm using path-based semantic annotations and reference ontologies for document transformations. Likewise, Nambi \textit{et al.} \cite{Nambi2014} proposed unification of knowledge through the development of a collection of ontologies to enable automatic service representation, composition, and discovery in dynamic IoT environments. However, the authors \cite{Kopke2017,Nambi2014} focus on modeling a set of ontologies, which describe devices and its functionalities. However, in an IoT network, all devices or users may not adopt the same set of ontologies for different contexts, which makes the interaction between them difficult \cite{Xiao2014}. As a solution of the mentioned problem, Xiao \textit{et al.} \cite{Xiao2014} proposed a user interoperability framework for seamless communication between IoT devices and users, which is based on the collaborative conceptualization theory. In their proposed framework, before communication, the device agents must build a collaborative semantically consistent sign of their devices to make a cosign dictionary. Thus, there is a need of automatic translator framework for semantic and syntactic interoperability to ensure seamless communication between devices and users without any prior information of the devices in terms of semantics and syntaxes. 

To address the above-mentioned issues, in this work, we present a middleware, named \textbf{M}iddleware for \textbf{S}emantic and \textbf{S}yntactic \textbf{I}nteroperability (MSSI), in PSF of IoT network. MSSI automatically translates user compatible syntaxes and semantics of the received information from the devices without prior information of devices' packet formats and semantic annotations. In summary, the \textit{contributions} of this paper are as follows.
\subsection{Contribution}\label{prob2:ssec:contribution}
\begin{itemize}
	\item We present a semantic interoperability framework to ensure seamless communication between users and devices using a multilayer perceptron (MLP).
	\item A novel algorithm is proposed for extracting raw and discriminative features, which are to be fitted to the MLP model as inputs to find out a correct standard name of the given attribute.
	\item In order to solve the syntactic disparities between users and devices, we propose a method of syntax translation of messages, which provides information in the user requested format, while presenting an algorithm for syntax identification and interoperability.
	\item To show the effectiveness of MSSI, we evaluate different parameters, while considering various used data formats and semantic annotations of attributes to ensure versatility of the proposed framework in a real-life scenario. 
\end{itemize}
The remainder of the paper is organized as follows. Section \ref{prob3:sec:related-works} summarizes the state-of-the-art existing works on seamless communication between the IoT devices and users in IoT. Section \ref{prob3:sec:system_architecture} presents the problem scenario and the proposed architecture of semantic and syntactic interoperability middleware. The process of semantic and syntactic translation is presented in Section \ref{prob3:sec:SeaCome_Framework}. Experimental setup and results of the proposed solution are shown in Section \ref{prob3:sec:performance_evaluation}. Finally, we conclude the paper and discuss future research directions of the work in Section \ref{prob3:sec:conclusion}.
\section{Related Works} \label{prob3:sec:related-works}
This section highlights the existing works related to seamless message sharing between IoT devices and users, while considering semantic and syntactic interoperability. K\"{o}pke \cite{Kopke2017} proposed a semantic translation method and semantic matching algorithm using path-based semantic annotations and reference ontologies for document transformations. Likewise, Nambi \textit{et al.} \cite{Nambi2014} proposed unification of knowledge through the development of a collection of ontologies to enable automatic service representation, composition, and discovery in dynamic IoT environments. Ontologies may be used to capture domain knowledge and infer new information from classes and relationships between them, which are not directly apparent from raw device data. 
On the other hand, to evaluate the interoperability of device, Henrik Dibowski proposed semantic evaluation interoperability model \cite{Dibowski2017}. Similarly, Yang \textit{et. al.} proposed interoperability framework between users and devices for semantic interoperability using divide-and-conquer technique \cite{Yang2019}. However, in an IoT network, all devices or users may not adopt the same set of ontologies for different contexts, which make the interaction between the devices and users difficult \cite{Xiao2014}. 

To solve the limitation of ontology, Xiao \textit{et al.} proposed a user interoperability framework for seamless communication between IoT devices and users, which is based on the collaborative conceptualization theory \cite{Xiao2014}. Moreover, to support the interaction between devices and users in different contexts, the authors adopted the collaborative \textit{sign} (cosign) dictionary approach. To make a cosign dictionary for all devices situated in different contexts, the device agents must collaborate to build semantic sign of all the devices. However, before communication, the device agents must collaboratively build a semantic consistent sign of their devices to make cosign dictionary. This induces the dependency of the device agents on the platform for registering their device.  

On the other hand, Moutinho \textit{et al.} \cite{Moutinho2018} addressed data association and added complementary data with the transmitter message, which is an issue of concern in semantic and syntactic interoperability. In \cite{Moutinho2018}, the proposed framework first collects the data format of both the sender and the receiver in the form of XML schema and verifies semantic compatibility. 
However, the framework is designed more specifically for supporting the association of temperature value with its unit and adding complementary data values in the transmitter data. Additionally, the authors only considered XML data format in their proposed scheme. 

In \cite{Kaed2018}, a semantic rule engine (SRE) is proposed to maintain interoperability between the IIoT network gateway and heterogeneous sensor/actuators communication. Semantic tags are used by the SRE to access the connected devices to ensure semantic interoperability between them. Similarly, Givehchi \textit{et al.}\cite{Givehchi2017} proposed a distinct interoperability layer between the cyber and the physical system to enable interoperability. A Common Information Model is applied in the interoperability layer based on the ISA95 industrial standard to ensure the interoperability for the legacy systems. The interoperability layer consists of three components --- raw data importer, mapper, and information provider, which collect the raw data from heterogeneous physical device and maps it to the receiver end format.

On the other hand, Ichise \cite{Ichise2008} used a learning based method for mapping between the concepts of ontologies based on multiple similarity measures. The measures defined in the paper are string-based, graph-based, instance based, and knowledge based. These measures are input to the Support Vector Machine to predict both positive and negative examples. Similarly, Doan \textit{et al.} \cite{Doan2002} adopted a unique approach towards the application of machine learning to solve the mapping problem on the Semantic Web. In their work, the authors described a way to find the most similar term for a concept in one ontology to a concept in another. The instances of concept A are used to learn a classifier for A which  classifies the instances of B, and vice versa. 

Synthesis: A critical analysis of the existing works unfold the existence of a research gap in semantic and syntactic interoperability in IoT network. Mainly, the existing works \cite{Kopke2017,Nambi2014,Dibowski2017} focused on semantic interoperability, which are inefficient in an IoT network using a set of ontologies. However, in the IoT network, all devices or users may not adopt the same set of ontologies for different contexts, which make the interaction between the devices and users difficult \cite{Xiao2014,Moutinho2018}. Some of them \cite{Xiao2014} attempted to solve the problem, but there is a dependency of the device agents on the platform for registering their device. In this paper, we present a framework, which automatically generates the users compatible syntax and semantics of the received information from the devices. 

\section{System Architecture} \label{prob3:sec:system_architecture}
\subsection{Problem Scenario}\label{prob3:ssec:problem_scenario}
The seamless communication between the devices and users is a big challenge in practice in the IoT scenario due to the usage of different data formats (syntaxes) and different semantic annotations to represent device data in any given context. The devices available in the market are manufactured by different vendors, who use different keywords and data formats such as JSON, XML, and CSV to represent the same concept. As an example, we present a problem scenario where a IoT device sends XML data packet of humidity sensor, as listed in Listing \ref{prob3:list:publisher_sent_xml}, and on the other hand, an user expects to receive data packet in JSON format containing the attributes date, time, humidity value, unit, device voltage, and battery voltage from the publisher in their compatible semantics, as listed in Listing \ref{prob3:list:subscriber_sent_json}. The publisher sends attributes named `DT', `T', `Humidity', `MeasuredUnit', `Dev\_Voltage', and `Batt\_Volt' instead of standard attribute names viz. date, time, humidity value, unit, device voltage, and battery voltage, respectively.
\begin{center}
	\begin{lstlisting}[caption={\footnotesize{Publisher sent XML message sample}}, label={prob3:list:publisher_sent_xml}]
	<ndata>
	<DT>29-Jan-17</DT>
	<T>23:30:06</T>
	<Humidity>14.48</Humidity>
	<MeasuredUnit>%</MeasuredUnit>
	<Dev_Voltage>4.97</Dev_Voltage>
	<Batt_Volt>12.73</Batt_Volt>
	</ndata>
	\end{lstlisting}
\vspace{-8mm}
\end{center}

\begin{center}
	\begin{lstlisting}[caption={\footnotesize{Subscriber expected JSON message sample}}, label={prob3:list:subscriber_sent_json}]
	{
	"MeasurementDate":"20-08-2015",
	"MeasurementTime":"22:50:05",
	"MeasurementRelativeHumidity":"93.64",
	"MeasurementUnit":"%",
	"MeasurementSystemVoltage":"4.98",
	"MeasurementBatteryVoltage":"12.57"
	}
	\end{lstlisting}
\end{center}

Similarly, the user uses attribute names `MeasurementDate', `MeasurementTime', `MeasurementRelativeHumidity', `MeasurementUnit', `MeasurementSystemVoltage', and `MeasurementBatteryVoltage' instead of standard attribute names date, time, humidity value, unit, device voltage, and battery voltage, respectively. In general, the meaning of usage attribute names are the same for both the users and the devices, but they use different keywords (semantics) to represent the same concept. Also, they use two different data formats to represent their data. Therefore, the usage of numerous semantics and several syntaxes by billions of the devices manufactured by different vendors, makes the seamless communication between the devices and users in IoT more complex.
\subsection{Proposed Architecture of MSSI}\label{prob3:ssec:proposed_architecture_SeaCom}
\begin{figure}[t]
	\centering
	\includegraphics[width=9cm]{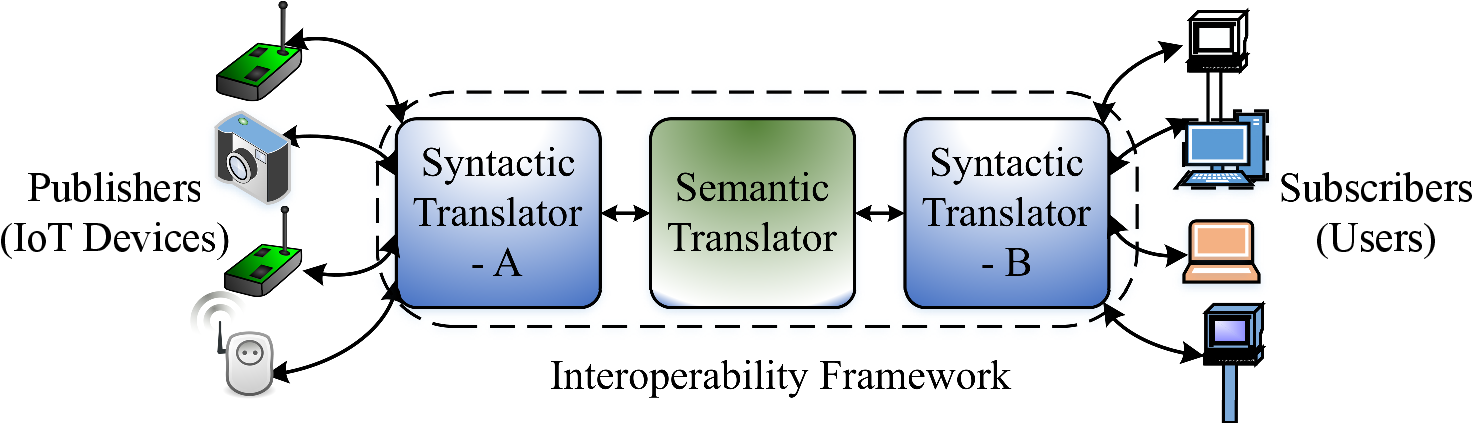}
	\caption{Proposed Architecture of MSSI}
	\label{prob3:fig:proposed_architecture_SeaCom}
\end{figure}

\begin{figure*}[t]
	\centering
	\includegraphics[width=14cm]{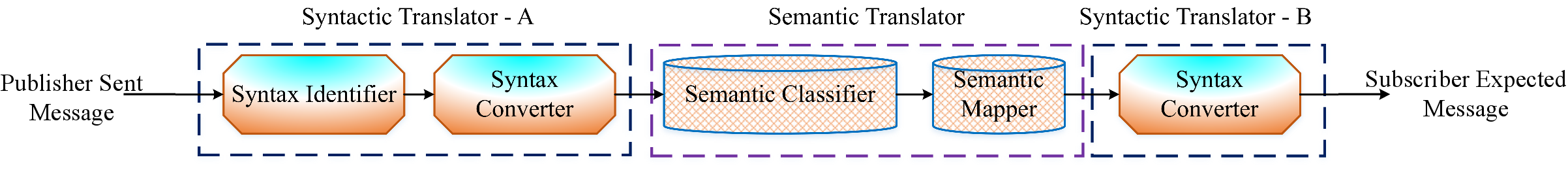}
	\caption{Functional block diagram of the proposed middleware for unified semantic and syntactic interoperability}
	\label{prob3:fig:syntactic_semantic_translator}
\end{figure*}

In this section, we propose an architecture of middleware framework to ensure unified semantic and syntactic interoperability between the users and the devices, as illustrated in Fig. \ref{prob3:fig:proposed_architecture_SeaCom}. The proposed framework consists of two parts --- a) syntactic translator and b) semantic translator. MSSI identifies the subscriber compatible syntaxes and semantics during the service request. The detailed process of syntax and semantic translation is shown in Fig. \ref{prob3:fig:syntactic_semantic_translator}. 
\subsubsection{Syntactic Translator}
The syntactic translator performs the syntax/data format translation of incoming message into a hierarchical data structure format and vice versa, as shown in Fig. \ref{prob3:fig:proposed_architecture_SeaCom}. In this paper, we only concentrate on the two extensively used information exchange formats --- JSON and XML, for syntactic interoperability, as the devices and users generally represent data using one of them data formats in the PSF. However, the method can be extended to more data formats as required. The syntactic translator is divided into two parts --- a) syntactic translator-A and b) syntactic translator-B, as shown in Fig. \ref{prob3:fig:proposed_architecture_SeaCom}. The devices send data using either the XML or JSON data formats to the MSSI middleware framework. The syntactic identifier of syntactic translator-A first identifies the actual data format of incoming data sent by publisher, then the syntax converter converts the data into a hierarchical data structure format. After semantic translation, the syntax converter of syntactic translator-B translates the data into the subscriber expected data format.
\subsubsection{Semantic Translator}\label{prob3:sssec:semantic_translator}
The semantic translator consists of two components called semantic classifier and semantic mapper. The devices use different attribute/element names to represent the same or different concept. So, the semantic translator first determines the standard meaning of each attribute of the received packet from the devices by the semantic classifier and then translates them into the user required attribute annotation. Each attribute has its own name and value. To identify the standard meaning of each attribute/element, we extract $26$ raw and $5$ discriminative features from the name and value of the attributes, respectively. Then we map these features to the MLP model as inputs and find out the standard name of the given attribute. Later, according to the shared semantic notation of users, the semantic mapper translates the standard attribute name to user's expected attribute name and appends the value of the corresponding attribute. The detailed process of semantic classification is discussed in Section \ref{prob3:ssec:semantic_translation}.
\section{MSSI: The Proposed Middleware for unified semantic and syntactic interoperability} \label{prob3:sec:SeaCome_Framework}
The detail of syntactic and semantic translations of incoming packets is shown in Fig. \ref{prob3:fig:syntactic_semantic_translator}.
\subsection{Syntactic Translation}\label{prob3:ssec:syntactic_translation}
In the first step, the proposed framework undertakes in order to ensure interoperability in IoT networks is to provide syntactic compatibility between the publishers and the subscribers of the messages. During the whole transformation process, in the flow between the publishers and the subscribers, data are syntactically translated twice. In the first step, translation of JSON or XML to a hierarchical data structure format takes place in order to provide easy access to data for semantic determination, as shown in Algorithm \ref{prob3:algo:syntax_translation_A}, where `isXML' function identifies that the incoming sensor packet is in XML format or not. The `isXML' function tries to identify standard syntax of a XML element, which has starting and ending tag. If the value of isXML function is false, the packet is in JSON format. After the identification process, the translator validates the incoming packet. During the second translation, intermediate representation to JSON or XML takes place in order to give data to the user in the subscriber requested data format, as shown in Algorithm \ref{prob3:algo:syntax_translation_B}.
\begin{algorithm}
	\scriptsize
	\caption{Syntax Translation: XML/JSON to a hierarchical data structure format}
	\begin{algorithmic}[1]
		\INPUT \State $\mathcal{F}_{m}$ \Comment{\texttt{$\mathcal{F}_{m}$ is the message received from the publisher}}
		\setcounter{ALG@line}{0}
		\OUTPUT \State $\mathcal{M}_{m}$ \Comment{\texttt{$\mathcal{M}_{m}$ is the hierarchical data structure format of the message}}
		\vspace{.5mm}
		\setcounter{ALG@line}{0}
		\PROC
		\If{isXML($\mathcal{F}_{m}$)} \Comment{\texttt{Checking the message containing XML data format or else}}
		   \If{isValideXML($\mathcal{F}_{m}$)} \Comment{\texttt{Validation of XML data format}}
		      \State $\mathcal{M}_{m}$= XML2Structure ($\mathcal{F}_{m}$) \Comment{\texttt{Convert XML to a hierarchical data structure format}}
		   \EndIf
		\Else \Comment{\texttt{Then the message containing JSON data format}}
		   \If{isValideJSON($\mathcal{F}_{m}$)}\Comment{\texttt{Validation of JSON data format}}
		     \State $\mathcal{M}_{m}$= JSON2Structure ($\mathcal{F}_{m}$) \Comment{\texttt{Convert JSON to a hierarchical data structure format}}
		   \EndIf
		\EndIf
		\Return $\mathcal{M}_{m}$
	\end{algorithmic}
	\label{prob3:algo:syntax_translation_A}
\end{algorithm} 
\begin{algorithm}
	\scriptsize
	\caption{Syntax Translation: a hierarchical data structure format to XML/JSON}
	\begin{algorithmic}[1]
		\INPUT \State $\mathcal{M}_{m}^{\dagger}$ \Comment{\texttt{$\mathcal{M}_{m}^{\dagger}$ is the hierarchical data structure format of the message after semantic translation}}
		\setcounter{ALG@line}{0}
		\OUTPUT \State $\mathcal{F}_{m}^{\dagger}$ \Comment{\texttt{$\mathcal{F}_{m}^{\dagger}$ is the message following subscriber data format}}
		\vspace{.5mm}
		\setcounter{ALG@line}{0}
		\PROC
		\If{isSubscriberFormat(XML)} \Comment{\texttt{Checking subscriber requested data format}}
		\State $\mathcal{F}_{m}^{\dagger}$= Structure2XML ($\mathcal{M}_{m}^{\dagger}$) \Comment{Convert a hierarchical data structure format to XML}
		\If{isValideXML($\mathcal{F}_{m}^{\dagger}$}) \Comment{\texttt{Validation of XML data format}}
		\Return $\mathcal{F}_{m}^{\dagger}$
		\EndIf
		\Else \Comment{\texttt{Then the message containing JSON data format}}
		\State $\mathcal{F}_{m}^{\dagger}$= Structure2JSON ($\mathcal{M}_{m}^{\dagger}$) \Comment{\texttt{Convert a hierarchical data structure format to JSON}}
		\If{isValideJSON($\mathcal{F}_{m}^{\dagger}$} \Comment{\texttt{Validation of JSON data format}}
		\Return $\mathcal{F}_{m}^{\dagger}$
		\EndIf
		\EndIf
	\end{algorithmic}
	\label{prob3:algo:syntax_translation_B}
\end{algorithm} 
\subsection{Semantic Translation}\label{prob3:ssec:semantic_translation}
In Section \ref{prob3:ssec:problem_scenario}, we presented a problem scenario and discussed how the usage of numerous semantics by billions of sensors, actuators, and systems manufactured by different vendors, make the seamless communication between the devices and users in IoT more complex. Similarly, the following example presents a scenario, where three publishers send XML messages containing data about relative humidity, ambient temperature, and soil moisture to the subscribers. The messages of three publishers are listed in Listings \ref{prob3:list:publisher_sample1},  \ref{prob3:list:publisher_sample2}, and \ref{prob3:list:publisher_sample3}. All three publishers use different semantic notation to represent the same or different concept. As an example, publishers $1$, $2$, and $3$ use `dat', `date', and `SensDate' to represent date, respectively and similarly for other attributes. From Listings \ref{prob3:list:publisher_sent_xml}, \ref{prob3:list:publisher_sample1}, \ref{prob3:list:publisher_sample2}, and \ref{prob3:list:publisher_sample3}, it is evident that the semantic annotations of attribute are extremely diverse. 

\begin{center}
	\scriptsize
	\begin{lstlisting}[caption={\footnotesize{Publisher 1 sent XML message of relative humidity sensor}}, label={prob3:list:publisher_sample1}]
	<Moto>
	<dat>2004-03-02</dat>
	<time>23:30:06.99</time>
	<packet_id>1</packet_id>
	<Node_id>8</Node_id>
	<humiditySensorValue>4.97</humiditySensorValue>
	<volt>2.68742</volt>
	</Moto>
	\end{lstlisting}
\end{center}
\begin{center}
	\scriptsize
	\begin{lstlisting}[caption={\footnotesize{Publisher 2 sent XML message of ambient temperature sensor}}, label={prob3:list:publisher_sample2}]
	<Motopacket>
	<date>2004-03-02</date>
	<T>23:30:06.99</T>
	<packet_No>776</packet_No>
	<motoid>34</motoid>
	<TempValue>14.97</TempValue>
	<sysvolt>2.68742</sysvolt>
	</Motopacket>
	\end{lstlisting}
\end{center}
\begin{center}
	\scriptsize
	\begin{lstlisting}[caption={\footnotesize{Publisher 3 sent XML message of soil moisture sensor}}, label={prob3:list:publisher_sample3}]
	<SM20Moto>
	<SensDate>1/1/2017</SensDate>
	<SensTime>23:30</SensTime>
	<MeasuredSensor>Soil Moisture</MeasuredSensor>
	<SoilmoistureValue>28.69</SoilmoistureValue>
	<InstalledDepth>20cm</InstalledDepth>
	</SM20Moto>
	\end{lstlisting}
\end{center}

Therefore, in the process of semantic translation, our first objective is to determine the standard meaning of each attribute/element of the incoming message from the devices. Once we find out the standard meaning of each attribute, it is easier to translate it into the user required attribute annotation. In this paper, we mainly focus on how to identify the standard meaning of the incoming attributes for semantic interoperability.

Every message contains multiple attributes and each attribute has its name and value. The identification process follows three steps: \textit{data organization}, \textit{feature extraction}, and \textit{attribute classification}.
\subsubsection{Data Organization}
Apart from different semantic annotations, different publishers also use special characters with the attribute name (or sometimes use a combination of upper and lower case letters), as listed in Listings \ref{prob3:list:publisher_sent_xml}, \ref{prob3:list:publisher_sample1}, \ref{prob3:list:publisher_sample2}, and \ref{prob3:list:publisher_sample3}. Therefore, we make a uniform text format to represent all attribute names and use the following preprocessing steps.
\begin{enumerate}[(i)]
	\item Remove special characters such as underscore (\_), hyphen (-), white space, and dot (.) from the attribute name.
	\item Convert all the attribute names to lower case.
	\item Find out the distinct possible attributes in the dataset. Let us consider $\mathcal{M}$ number of distinct possible attributes or classes in the dataset. The set of possible distinct output classes is defined as,	$\mathcal{C}_{S}=\{\mathcal{C}_{1}, \mathcal{C}_{2},\mathcal{C}_{3}, \dots, \mathcal{C}_{\mathcal{M}}\}$. Example: date, time, sensor name, unit, device voltage, battery voltage, network id, device address, and soil depth.
	\item Find out the possible name of each distinct attribute. Let us consider that there exists $\mathcal{N}$ number of possible names for $\mathcal{C}_{i}$ distinct class. The set of possible names in $\mathcal{C}_{i}$ distinct class is expressed as, $\mathcal{C}_{i}=\{\mathcal{C}_{1}^{i}, \mathcal{C}_{2}^{i},\mathcal{C}_{3}^{i}, \dots, \mathcal{C}_{\mathcal{N}}^{i}\} \quad \forall i\in\mathcal{M}$. For example, a distinct attribute/class `date' different possible names can be dat, d, sensingdate, measureddate, period, dt, measurementdate, and sensdat
\end{enumerate}
\begin{figure}[t!]
	\centering
	\subfigure[Raw features vector]
	{
		\includegraphics[width=8.5cm]{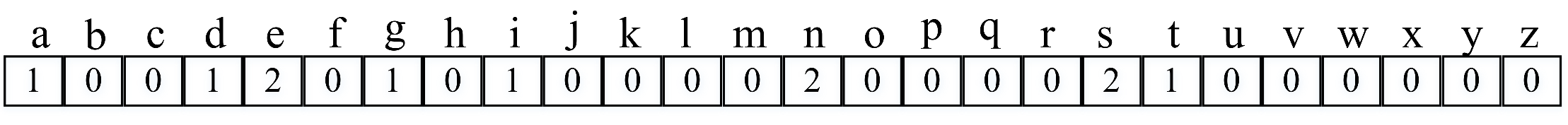}
		\label{prob3:fig:example_raw_feature_extraction}	
	}
	\subfigure[Discriminative features vector]
	{
		\includegraphics[width=7cm]{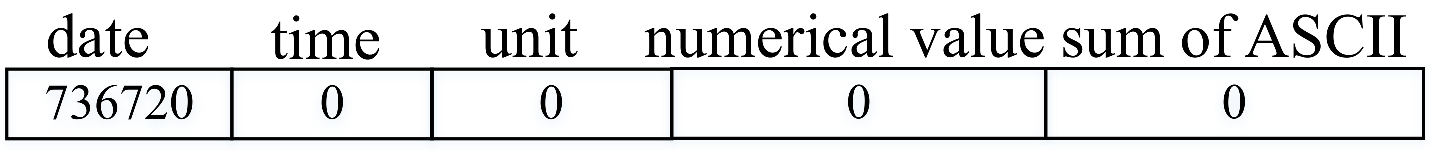}
		\label{prob3:fig:example_discrimative_feature_extraction}	
	}
	\caption{Example of feature extraction}
	\label{prob3:fig:example_feature_extraction}
\end{figure}
\subsubsection{Feature Extraction}\label{prob3:sssec:feature_extraction}
In this process, our objective is to identify unique features, which are to be fitted to the MLP model as inputs to find out a correct standard name of the given attribute. The name of the attribute consists of only alphabets or alphanumerics. On the other hand, the value of the attribute is represented by alphabets, alphanumerics, numerics, numerics with special character, only special characters, or alphabets with special character. The feature extraction process is divided into two categories --- \textit{raw feature extraction} and \textit{discriminative feature extraction}.

\textit{a) Raw Feature Extraction:} Raw features are defined as the frequency of occurrence of each alphabet/letter in the attribute name. English alphabet consists of $26$ letters. According to the frequency of occurrence of each letter, we extract the $26$-dimension feature vector $\mathcal{F}_{\mathcal{R}}$ from the name of the attribute. The process of raw feature extraction from the attribute name is presented in Algorithm \ref{prob3:algo:feature_extraction}, where the function letterCount() is used to calculate the frequency of occurrence of each letter in the attribute name.

\textit{b) Discriminative Feature:} We additionally extract $5$-dimensional discriminative features to preserve intraclass distance and increase interclass distance. Towards that, we find out the distinct pattern of the value of all attributes. Accordingly, the discriminative features are date, time, unit, numerical value, and sum of ASCII of all characters, which are extracted from the value of the attribute. Algorithm \ref{prob3:algo:feature_extraction} presents the process of extraction of discriminative features from the attribute value, where $\mathcal{F}_{\mathcal{D}}$ is the $5$-dimension discriminative feature vector. 
\begin{center}
	\begin{lstlisting}[caption={\footnotesize{Attribute Annotation}}, label={prob3:list:feature_extaction_example}]
	<sensingdate>2004-03-02</sensingdate>
	\end{lstlisting}
\end{center}
\begin{algorithm}
\scriptsize
	\caption{Features Extraction}
	\begin{algorithmic}[1]
		\INPUT \State $\mathcal{A}_{n}$, $\mathcal{A}_{v}$, $\mathcal{F}_{\mathcal{R}(26X1)}$, $\mathcal{F}_{\mathcal{D}(5X1)}$ \Comment{\texttt{$\mathcal{A}_{n}$, $\mathcal{A}_{v}$, $\mathcal{F}_{\mathcal{R}}$, and $\mathcal{F}_{\mathcal{D}}$ are the attribute name, attribute value, zeros matrix with $26$ dimension, and zeros matrix with $5$ dimension, respectively}}
		\setcounter{ALG@line}{0}
		\OUTPUT \State $\mathcal{F}_{total}$ \Comment{\texttt{$31$ dimension feature vector}}
		\vspace{.5mm}
		\setcounter{ALG@line}{0}
		\PROC
		\State $\mathcal{F}_{\mathcal{R}}$[:,1] = letterCount($\mathcal{A}_{n}$) \Comment{\texttt{Frequency of occurrence of each letter in the attribute name}}
		\If{isNum($\mathcal{A}_{v}$)} \Comment{\texttt{Checking the attribute value is only numeric or else}}
			\State $\mathcal{F}_{\mathcal{D}}$[4,1]=$\mathcal{A}_{v}$ \Comment{\texttt{Copy the numeric value in index $4$ of $5$ dimensional discriminative feature vector}}
		\Else
			\If{isCharThere($\mathcal{A}_{v}$,`-') $\mid$ isCharThere($\mathcal{A}_{v}$,`/') \& charCount($\mathcal{A}_{v}$,`-')$>$$1$$ \mid$ charCount($\mathcal{A}_{v}$,`/')$>$$1$} \Comment{\texttt{To find out date format}}
				\State $\mathcal{F}_{\mathcal{D}}$[1,1]=date2num ($\mathcal{A}_{v}$) \Comment{\texttt{Convert date to numerical}}
			\ElsIf{isCharThere($\mathcal{A}_{v}$,`:')} \Comment{\texttt{To find out time format}}
				\State $\mathcal{F}_{\mathcal{D}}$[2,1]=ASCII($\mathcal{A}_{v}$) \Comment{\texttt{Convert time to ASCII of character `:'}}
			\ElsIf{isCharThere($\mathcal{A}_{v}$,`\%')} \Comment{\texttt{To find out special character for unit}}
				\State $\mathcal{F}_{\mathcal{D}}$[3,1]=ASCII($\mathcal{A}_{v}$)\Comment{\texttt{Convert `\%' to ASCII value}}
			\Else 
				\State temp = removeSpeChar($\mathcal{A}_{v}$,`\_',`-',` ',`.')\Comment{\texttt{Remove special characters from text}}
				\State temp = lowerCase(temp)\Comment{\texttt{Converting to lower case}}
				\State $\mathcal{F}_{\mathcal{D}}$[5,1]=sum(ASCII(temp))\Comment{\texttt{Calculating sum of ASCII value of all character in the text}}
			\EndIf
		\EndIf 
		\State $\mathcal{F}_{total}$=[$\mathcal{F}_{\mathcal{R}}$ ; $\mathcal{F}_{\mathcal{D}}$]\Comment{\texttt{Making $31$ dimension features vector}}\\
		\Return $\mathcal{F}_{total}$ 
	\end{algorithmic}
	\label{prob3:algo:feature_extraction}
\end{algorithm} 

We present an example of an attribute in Listing \ref{prob3:list:feature_extaction_example}. Using Algorithm \ref{prob3:algo:feature_extraction}, the calculated raw and discriminative features of the attribute are shown in Figs. \ref{prob3:fig:example_raw_feature_extraction} and \ref{prob3:fig:example_discrimative_feature_extraction}. The combination of raw and discriminative features of the attribute is the $31$ dimension feature vector, which is to be mapped to the MLP model as inputs to find out the standard name of the given attribute.

\begin{figure}[t!]
	\centering
	\subfigure[Date]
	{
		\includegraphics[width=8.5cm]{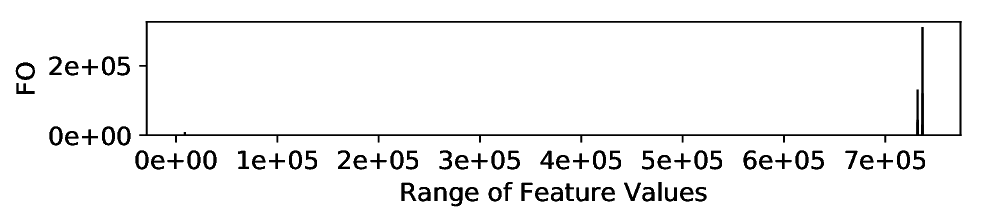}
		\label{prob3:fig:histogram_discriminative_features_date}	
	}\vspace{-2mm}
	\subfigure[Time]
	{
		\includegraphics[width=8.5cm]{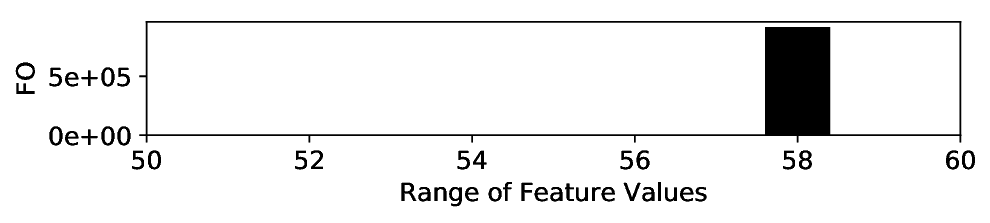}
		\label{prob3:fig:histogram_discriminative_features_time}	
	}\vspace{-2mm}
	\subfigure[Unit]
	{
		\includegraphics[width=8.5cm]{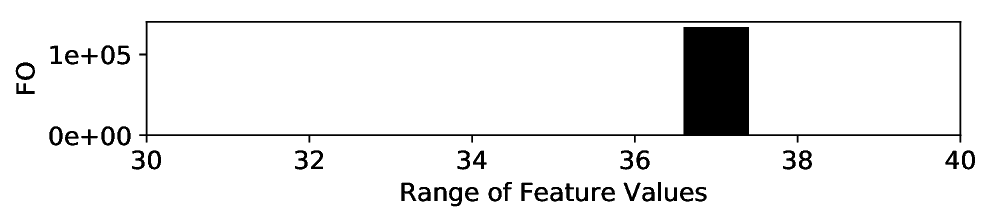}
		\label{prob3:fig:histogram_discriminative_features_unit}	
	}\vspace{-2mm}
	\subfigure[Numerical value]
	{
		\includegraphics[width=8.5cm]{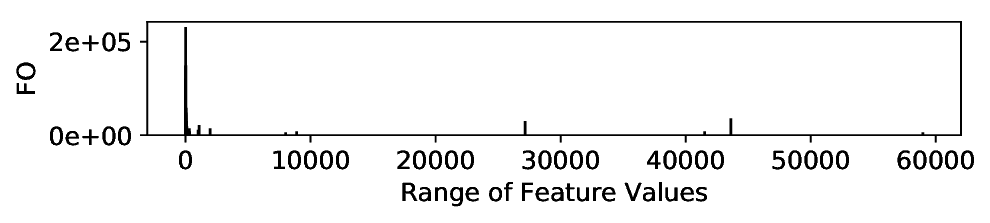}
		\label{prob3:fig:histogram_discriminative_features_numerical}	
	}\vspace{-2mm}
	\subfigure[Sum of ASCII of characters]
	{
		\includegraphics[width=8.5cm]{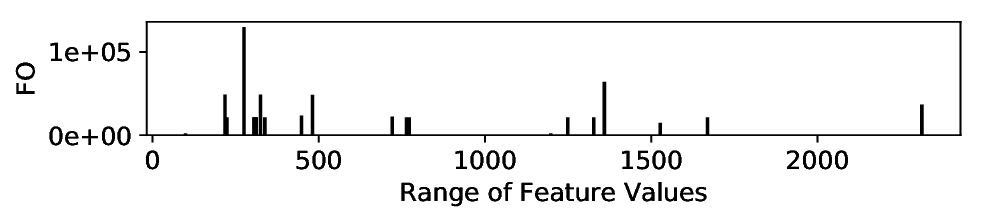}
		\label{prob3:fig:histogram_discriminative_features_sumofAcii}	
	}
	\caption{Histogram of discriminative features}
   \label{prob3:fig:histogram_discriminative_features}
\end{figure}

The histogram of discriminative features is shown in Fig. \ref{prob3:fig:histogram_discriminative_features}. Figs. \ref{prob3:fig:histogram_discriminative_features_date}, \ref{prob3:fig:histogram_discriminative_features_time}, \ref{prob3:fig:histogram_discriminative_features_unit}, \ref{prob3:fig:histogram_discriminative_features_numerical}, and \ref{prob3:fig:histogram_discriminative_features_sumofAcii} show the range of date, time, unit, numerical value, and sum of ASCII of string, respectively. From these figures, it is evident that the range of each discriminative feature values is distinct from the other features. In Fig. \ref{prob3:fig:histogram_discriminative_features_date}, the value of the date feature lies in the order of $10^5$. Similarly, the value feature belongs from $0$ to $10^4$ order, as shown in Fig. \ref{prob3:fig:histogram_discriminative_features_time}. On the other hand, the range of time and unit features is $57.5$ to $58.5$ and $36.5$ to $37.5$, respectively. However, there are some overlapping features between two features, but that can be solved by using raw features. Therefore, the discriminative features increase the interclass distance and reduce the intraclass distance, which help to classify all attributes properly. The discriminative features extraction is the \textit{core contribution} of this work.
\subsubsection{Attribute Classification}\label{prob3:sssec:attribute_classification}
In order to identify the standard name of each attribute, we use the Gaussian Mixture Model (GMM), Naive Bayes (NB), and the Artificial Neural Network (ANN) methods. The details derivation of GMM is described in Appendix \ref{appendix_GMM} \cite{Bishop2006}. In this experiment, the Expectation Maximization (EM) algorithm is used to maximize the value of log likelihood of each class for GMM \cite{Evangelidis2018}. The derivation of NB is described in Appendix \ref{appendix_NB} \cite{John1995}. On the other hand, we use multilayer perceptron (MLP) algorithm for ANN and the derivation of MLP is described in Appendix \ref{appendix_MLP} \cite{Bishop2006}. The proposed neural network has three layers, i.e., input, hidden, and output, while considering different parameters such as learning rate, momentum, and epoch. Furthermore, random numbers are used for setting the initial weights of the connections between nodes, and also for shuffling the training data. Error backpropagation algorithm is used to iteratively update the parameters of MLP \cite{Tisan2016}. As an activation function, we use sigmoid function. The extracted $31$ features vectors of an attribute, as discussed in Section \ref{prob3:sssec:feature_extraction}, are mapped to these classifiers as inputs to find out the standard name of the given attribute, which is produced as an output.
\section{Performance Evaluation} \label{prob3:sec:performance_evaluation}
\subsection{Experimental Setup} \label{prob3:ssec:experimental_setting}
In our experiment, we used public datasets \cite{MIT2023,Murdoch2023} as well as a private dataset \cite{Roy2020} of various sensor data, while considering weather and agriculture sensor parameters such as ambient temperature, relative humidity, solar radiation, wind speed, rainfall, soil moisture, air pressure, luminosity, and soil temperature. It is noteworthy that all the sources have their own attribute annotation and its corresponding value. Also, the data in these datasets have no data format. Thus, to use these datasets in our experiment, we converted all excel data into JSON/XML data format. As an example, the conversion of sensor data into XML/JSON format is listed in Listings \ref{prob3:list:publisher_sent_xml}, \ref{prob3:list:subscriber_sent_json}, \ref{prob3:list:publisher_sample1},  \ref{prob3:list:publisher_sample2}, and \ref{prob3:list:publisher_sample3}.

Apart from the above-mentioned semantic annotations to ensure the robustness and adaptability of the proposed framework, we experimented with other possible semantic annotations. To do this, we used the semantic annotations defined in \cite{IoTIgnite2023,SNON2023,Moutinho2018}, where the authors have used various different annotations. For example, to define a device id ($\mathcal{C}_9$), the authors used nodeid, moteid, iotdeviceid, deviceid etc. Similarly, we followed the semantic annotations for other classes also. In the dataset, there are $15$ ($\mathcal{M}$) distinct possible attributes/classes such as date ($\mathcal{C}_1$), time ($\mathcal{C}_2$), sensor name ($\mathcal{C}_3$), sensor value ($\mathcal{C}_4$), unit ($\mathcal{C}_5$), device voltage ($\mathcal{C}_6$), battery voltage ($\mathcal{C}_7$), network id ($\mathcal{C}_8$), device id ($\mathcal{C}_9$), channel id ($\mathcal{C}_{10}$), momentary ($\mathcal{C}_{11}$), automatic readout ($\mathcal{C}_{12}$), epoch ($\mathcal{C}_{13}$), soil depth ($\mathcal{C}_{14}$), and description ($\mathcal{C}_{15}$). Therefore, the number of input feature vectors and classes are $31$ and $15$, respectively.  

The experimental setup is discussed in Table \ref{prob3:table:sexperimental_setup}. In the datasets, total number of messages for training and test is $2633597$ and $754183$, respectively. The number of attributes in a message varies from source-to-source. However, the total number of attributes for training and testing is $4961876$ and $1158412$, respectively. It is noteworthy that the proposed solution approach consists of two parts --- syntactic interoperability and semantic interoperability. The solution for syntactic interoperability of the messages is presented in Section \ref{prob3:ssec:syntactic_translation}. For semantic interoperability, after feature extraction, our objective is to determine the standard meaning of each attribute/element of the incoming message from the devices using GMM, NB, and MLP, as discussed in Section \ref{prob3:sssec:attribute_classification}. Then, it is easier to map the standard attribute name into the user required attribute annotation by the semantic mapper, discussed in Section \ref{prob3:sssec:semantic_translator}. 
\begin{table}[t]
	\scriptsize
	\caption{Experimental setup}
	\label{prob3:table:sexperimental_setup}
	\begin{center}
		\begin{tabular}{ | l | l | }
			\hline
			\textbf{Parameter}                                      & \textbf{Value}               \\ \hline
			Number of Gaussians                                     & 2-16                         \\ \hline
			Learning rate for MLP                                          & 0.1                          \\ \hline
			Momentum for MLP                                              & 0.2                          \\ \hline
			Epoch for MLP                                                  & 500                          \\ \hline
			Neurons in input layer                                  & 31                           \\ \hline
			Number of hidden layer                                  & 1                            \\ \hline
			Number of neurons in the hidden layer                   & 20-35                        \\ \hline
			Neurons in output layer                                 & 15                           \\ \hline
			Input features vector including class label & 32 \\ \hline
			Output classes & 15 \\ \hline
			Total number of training data set & 4961876 \\ \hline
			Total number of test data set    & 1158412 \\ \hline
		\end{tabular}
	\end{center}
\end{table}
 \begin{figure}[t]
	\centering
	\includegraphics[width=7cm]{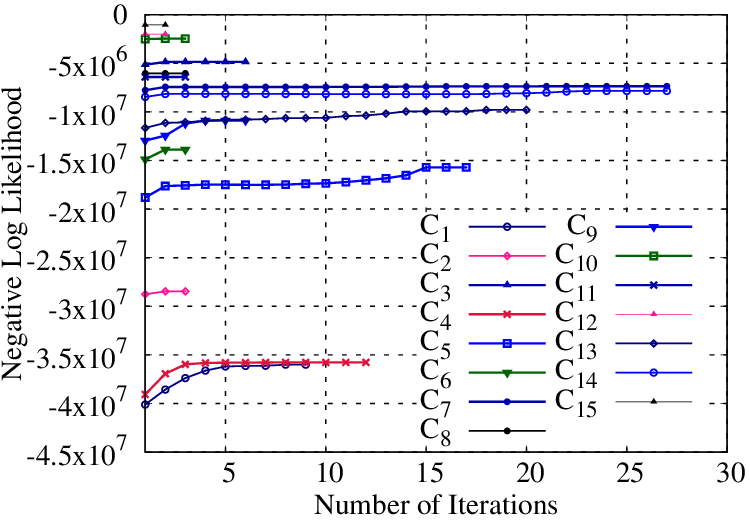}
	\caption{Negative log likelihood of different classes}
	\label{prob3:plot:negative_log_likelihood}
\end{figure}
 \subsection{Results and Discussion} \label{prob3:ssec:result_discussion}
 This section presents the performance of MSSI to show the effectiveness of the proposed solution.
\subsubsection{Convergence Analysis}
We evaluate the negative log likelihood (NLL) of all classes for training dataset in GMM, as shown in Fig. \ref{prob3:plot:negative_log_likelihood}. The value of NLL acts as a stopping criteria to preserve the optimal parameters of GMM viz. mean, variance, and \textit{a priori} weights. We stop GMM training if the difference between the present and the previous NLL value is less than $10^{-3}$, which is negligible. From Fig. \ref{prob3:plot:negative_log_likelihood}, it is evident that all the classes properly converged after a certain number of iterations. In addition, the value of negative log likelihood for each class is distinct from other classes. Hence, the distinct value of negative log likelihood for each class helps to calculate unique value of mean and variance of each class. Therefore, the distinct values of negative log likelihood for all classes improve the accuracy of attributes classification. 
\begin{figure}[h!]
	\centering
	\subfigure[Varying Gaussians]
	{
		\includegraphics[width=4.2cm]{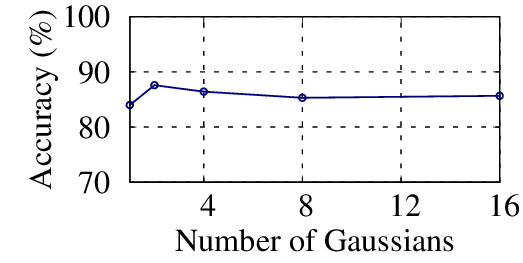}
		\label{prob3:plot:GMM_accuracy}	
	}
	\subfigure[Varying Neurons]
	{
		\includegraphics[width=4.2cm]{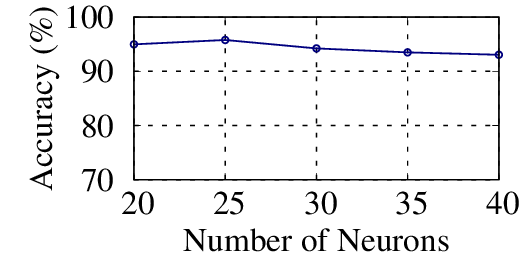}
		\label{prob3:plot:accuracy_ANN}	
	}
	\caption{Accuracy analysis. (a) Varying Gaussians. (b) Varying Neurons.}
	\label{prob3:fig:accuracy_analysis_varying_parameter}
\end{figure}	
\subsubsection{Accuracy Analysis}
We analyze the classification accuracy of attributes using GMM and MLP to choose the optimal value of different parameters of these classifiers to get the highest classification accuracy for the test dataset, as shown in Fig. \ref{prob3:fig:accuracy_analysis_varying_parameter}.

\textit{a) Varying Gaussians:} Fig. \ref{prob3:plot:GMM_accuracy} presents the classification accuracy of attributes using GMM, while varying the number of Gaussians ($\mathcal{N}_{G}$). $\mathcal{N}_{G}$ is chosen empirically to find out the optimal number of Gaussians for our experimental test dataset. From Fig. \ref{prob3:plot:GMM_accuracy}, it is shown that the classification accuracy is maximum when $\mathcal{N}_{G}$ is $2$ for GMM model. The classification accuracy of one Gaussian is less than two Gaussians because the data distribution of the proposed features is not properly captured by one Gaussian. On the other hand, the classification accuracy of more than two Gaussians is less than two Gaussians due to the problem of data insufficiency problem, which may occur for increasing $\mathcal{N}_{G}$. Furthermore, when $\mathcal{N}_{G}$ increases, GMM tries to form overlapping clusters, which is unnecessary. Additionally, the computation of additional means, variance, and the weights become more expensive. Thus, Fig. \ref{prob3:plot:GMM_accuracy} signifies that two Gaussians are sufficient to capture the data distribution of the proposed features. Therefore, $\mathcal{N}_{G}$ is set to $2$ for other experiments in the rest of the paper.
	
\textit{b) Varying Neurons/Nodes:} Similar to GMM, we conducted another experiment to select optimal number of neurons ($\mathcal{N}_N$) of hidden layer $1$ for our test dataset. Fig. \ref{prob3:plot:accuracy_ANN} shows the classification accuracy using the MLP model, while varying the number of neurons. $\mathcal{N}_N$ of hidden layer $1$ is chosen to find out the optimal number for our test dataset. From this figure, it is evident that the classification accuracy is maximum when $\mathcal{N}_N$ is $25$ for the MLP model. Therefore, the standard value of $\mathcal{N}_N$ of hidden layer $1$ is chosen as $25$, which is to be used for other experiments in the rest of the paper. The choice of $1$ hidden layer projects the non-separable input data into a high dimensional space, where the patterns are expected to be linearly separable. Once the input features are separable at hidden layer $1$, the additional hidden layers are not necessary, which may also become computationally expensive to calculate the weights between the additional layers.
\begin{figure}[h!]
	\centering
	\subfigure[Using GMM]
	{
		\includegraphics[scale=.38]{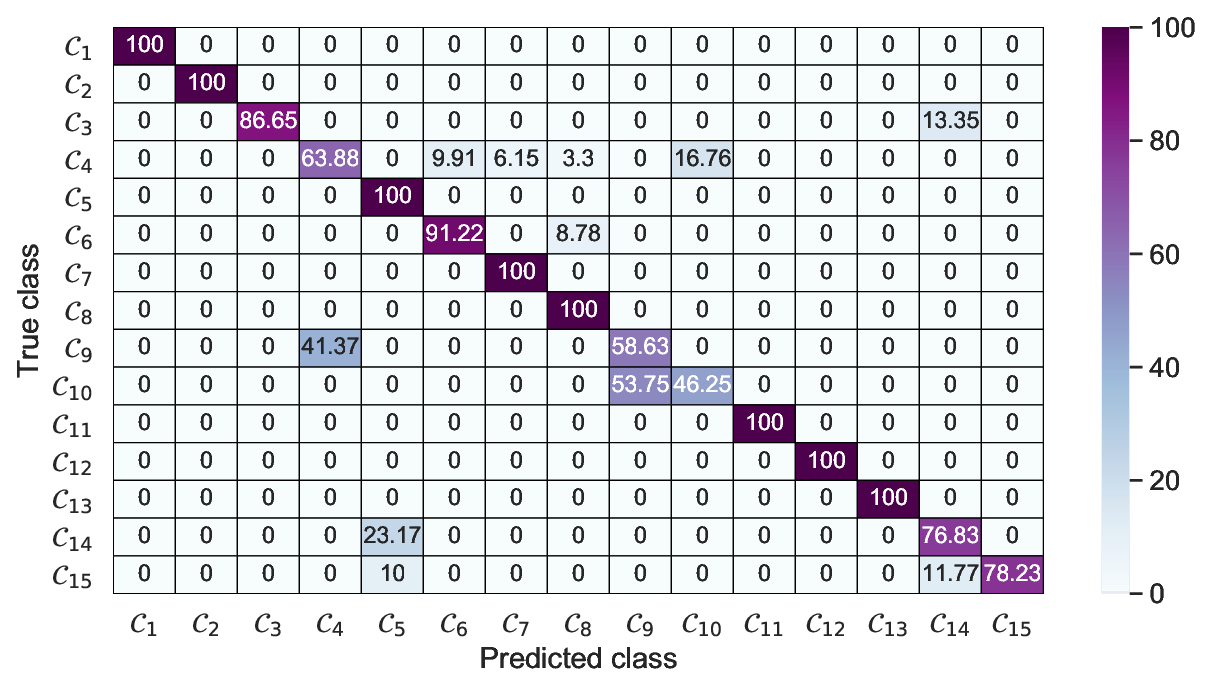}
		\label{prob3:plot:GMM_confusoion_matrix}	
	}
	\subfigure[Using Naive Bayes]
	{ 
		\includegraphics[scale=.38]{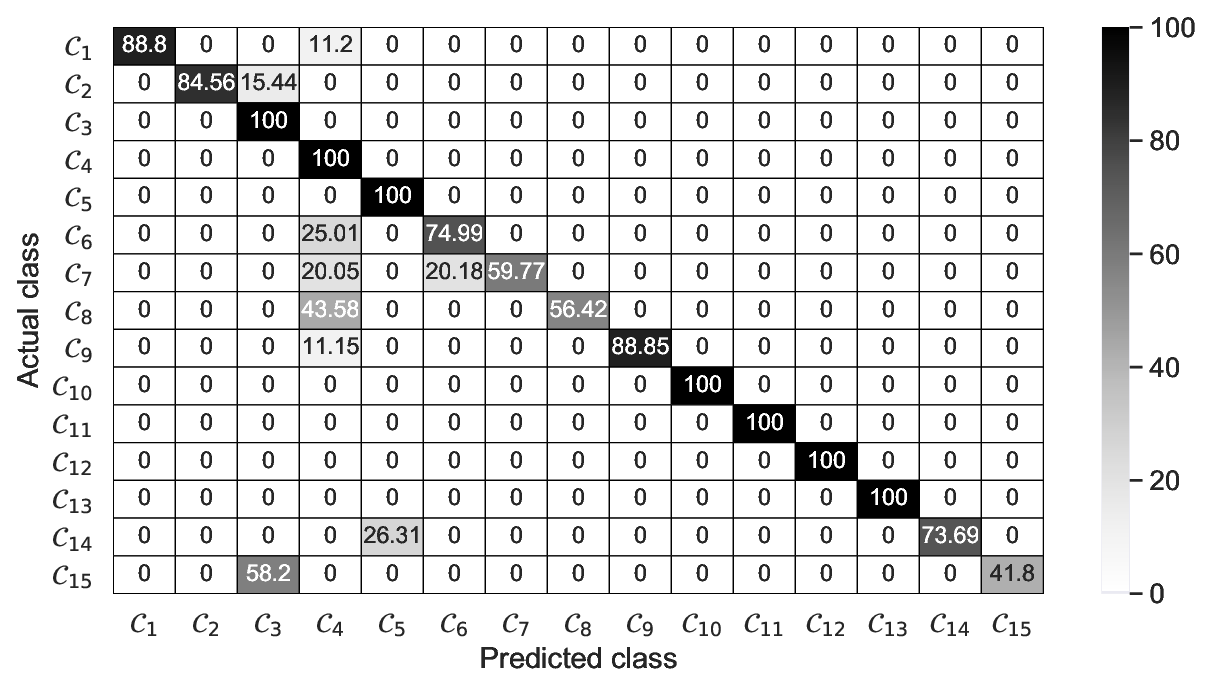}
		\label{prob3:plot:Naive_confusoion_matrix}	
	}
	\subfigure[Using MLP]
	{ 
		\includegraphics[scale=.38]{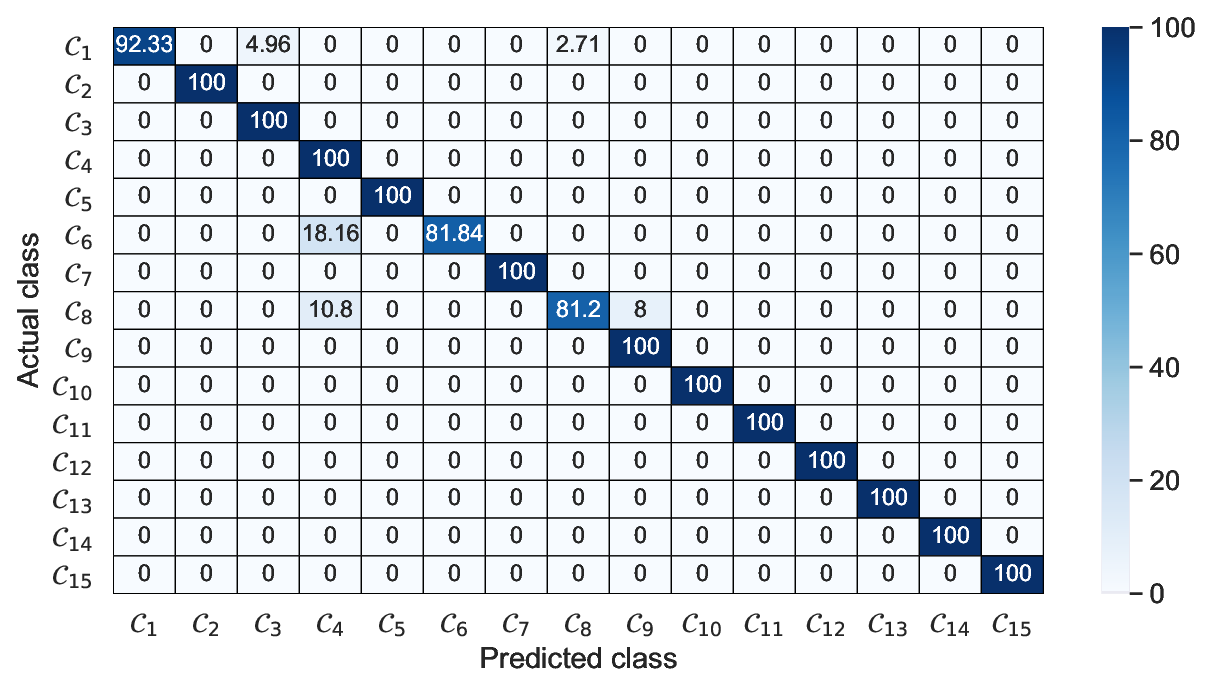}
		\label{prob3:plot:ANN_confusoion_matrix}	
	}
	\caption{Confusion matrix. (a) Using GMM. (b) Using Naive Bayes. (c) Using MLP.}
	\label{prob3:fig:aconfusion_matrix}
\end{figure}
\subsubsection{Confusion Matrix Analysis} 
We evaluate the confusion matrix of this attribute classification problem using GMM, as shown in Fig. \ref{prob3:plot:GMM_confusoion_matrix}. Fig. \ref{prob3:plot:GMM_confusoion_matrix} shows a summary of the percentage of correct and incorrect predictions of each class using GMM. From this figure, it is evident that all the classes are properly predicted on the test dataset. However, the classification accuracy of some classes such as $\mathcal{C}_{4}$, $\mathcal{C}_{9}$, $\mathcal{C}_{10}$, $\mathcal{C}_{14}$, and $\mathcal{C}_{15}$ is lesser compared to the other classes. We observe that the misclassification of a class occurs due to the closeness of mean vectors of the true positive class and false negative classes in GMM. 

Similarly, Fig. \ref{prob3:plot:Naive_confusoion_matrix} is evident that all the classes are properly predicted on the test dataset using NB. However, the classification accuracy of some classes such as $\mathcal{C}_{7}$, $\mathcal{C}_{8}$, and $\mathcal{C}_{15}$ is lesser compared to the other classes. We observe that the misclassification of a class occurs due to the low values of precision, recall, and f-measure of these classes. 

Fig. \ref{prob3:plot:ANN_confusoion_matrix} presents a summary of the percentage of correct and incorrect predictions of each class using MLP. Fig. \ref{prob3:plot:ANN_confusoion_matrix} shows that all the classes are properly classified with a higher rate of true positive (TP). Although the rate of TP of class $\mathcal{C}_{6}$ and $\mathcal{C}_{8}$ is slightly lower than that of other classes using MLP. However, the overall rate of TP of all classes using MLP is significantly higher compared to that using GMM and NB due to the very high average value of precision, recall, and f-measure of all classes in the MLP model, which is $0.955$, $0.958$, and $0.954$, respectively.
\subsubsection{Accuracy of Semantic Classification} The overall accuracy of semantic classification using GMM, NB, and MLP is $87.58$\%, $87.27$\%, and $95.78$\%, respectively, as shown in Table \ref{prob3:table:classification_compaision}. It is found that the MLP algorithm provides more accuracy compared to GMM and NB, as MLP generates the posterior vector, where we can obtain the probability of the test features belonging to other classes as well. On the other hand, in the GMM, the test data is evaluated based on the log likelihood score generated from the GMM parameters. The GMM parameters do not provide the probability of the test features belonging to other classes, which only give the likelihood score. On the other side, NB has a higher error rate with respect to the growth of the size of data-instances. Thus, MLP is adopted to determine the standard meaning of each attribute/element of the incoming message from the devices to solve semantic interoperability problem in IoT network. The overall classification accuracy is $95.78$\%.
\begin{table}[t]
	\scriptsize
	\caption{Classification accuracy comparison}
	\label{prob3:table:classification_compaision}
\begin{center}
	\begin{tabular}{|c|c|}
		\hline
		\textbf{Name of classifier} & \textbf{Percentage of accuracy} \\ \hline
		GMM                         & $87.58$                         \\ \hline
		Naive Bayes                 & $87.27$                         \\ \hline
		MLP                         & $95.78$                         \\ \hline
	\end{tabular}
\end{center}
\end{table}
\subsection{Real-Life Applicability of MSSI}
In this section, we discuss the practical use of our proposed MSSI in a smart industrial scenario. Typically in an IoT based smart factory environment, the manufacturing equipments are equipped with sensors, which collect data and forward it to the sensor-cloud servers for further analysis and critical observation. This sensor data is observed by various units of the factory such as manufacturing unit, maintenance unit, inventory tracking unit and so on. Sensor-cloud environment virtually connects these units with the physical sensors and fulfills the different demands generated by various units. As an example, manufacturing unit requests sensor-cloud for the consistent real-time sensor data so that if there are any intricacies during the manufacturing process, they will be detected beforehand. But, the sensors use different semantics and data formats to represent the data as these equipments are manufactured by different vendors. Also, each unit use its own semantic annotation and data formats. Hence, the communication between physical sensors and the units is not possible. Therefore, there is a need to have an adaptive semantic and syntactic interoperability framework installed in sensor-cloud for seamless communication between the sensors and the units.
\section{Conclusion}\label{prob3:sec:conclusion} 
In this paper, we propose a middleware framework, which has the capability to automatically translate the subscribers compatible syntax and semantics of the received message from the publishers without prior knowledge of publishers semantics and syntax in PSF of IoT network. We propose a method of syntax translation of messages to solve the syntactic disparities between publishers and subscribers. An MLP-based semantic interoperability framework is proposed to translate device information to the user requested semantics. Additionally, an algorithm is proposed for extracting raw and discriminative features, which are to be fitted to the GMM, NB, and MLP as inputs. To show the effectiveness of MSSI, we evaluated different parameters, while considering various used data formats and semantics annotations of attributes to ensure versatility of the proposed framework in the practical scenario. The overall classification accuracy using MLP is $95.78$\% for determining the standard meaning of each attribute of the incoming message from the publisher to solve the semantic interoperability problem in an IoT network.

To achieve seamless communication, there is a need of communication protocols interoperability apart from semantic and syntactic interoperability. The proposed framework can be extended in the future to address this issue. 

\appendices
\section{}
A classification function can be represented as below,
\begin{equation}
	\mathcal{Y}=f(\mathcal{X})
\end{equation}
$\mathcal{X} \in \mathbb{R}^{N\times\mathcal{F}}$ is the input features vector, where $N$ and $\mathcal{F}$ are the number of samples and  the dimension of input feature vectors, respectively. On the other hand, $\mathcal{Y} \in \mathbb{R}^{\mathcal{N}\times\mathcal{M}}$ is the classification output (also called the ``class"), where $\mathcal{M}$ is the number of classes in the dataset. $y$ is an example of $M$ dimensional output, which is one-hot encoding vector.
\subsection{Gaussian Mixture Model (GMM)}\label{appendix_GMM}
GMM is a type of generative classification, which is a probabilistic model of the probability density distribution of features. The goal here is to learn the clusters present in the data in an unsupervised manner as per the equation given below. The same approach is further tailored to support classification, where the number of clusters learnt from the training data is equal to the no of classes in classification and thus every cluster represents a class in the data.  
\begin{equation}
	\mathcal{P}(\textbf{x})=\sum_{k=1}^{\mathcal{M}}\mathcal{P}(x|y_k)\mathcal{P}(y_k)
\end{equation}
where, $\mathcal{P}(y_k)$ is uniform distribution. The typical assumption is that $\mathcal{P}(\textbf{x}|y_k)$ follows normal distribution, which represents each class in the data. Therefore, $\mathcal{P}(\textbf{x}|y_k)$ can be written as
\begin{equation}
	\mathcal{P}(\textbf{x}|y_k)=\mathcal{N}(\textbf{x} | \mu_k, \sigma_k^{2})
\end{equation}
where $\mu_k$ and $\sigma_k^{2}$ are the mean and variance of the $k_th$ class. We use Expectation maximization (EM) algorithm to predict the learning parameters such as mean ($\mu$) and variance ($\sigma^{2}$). 

Once the parameters of the distribution for each class is found out from the data, it can then be used for evaluating the class of a data point at test time.

In case of testing of new input $\hat{\textbf{x}}$, the classification output is represented as
\begin{equation}
	\hat{y}=\arg\max_{k \in {1......\mathcal{M}}}\mathcal{P}(\hat{\textbf{x}}|y_k)=\mathcal{N}(\hat{\textbf{x}} | \mu_k, \sigma_k^{2})
\end{equation}
where $\mu_k$ and $\sigma_k^{2}$ are used to classify new input $\hat{\textbf{x}}$.

\subsection{Naive Bayes (NB)}\label{appendix_NB}
Naive Bayes is a conditional probability model, where the function $f(\mathcal{X})$ predicts highest conditional probability of given inputs.

The instance probability of each class is
\begin{equation}
	\mathcal{P}(y_k | x_1, x_2,.......,x_{\mathcal{F}})=\frac{\mathcal{P}(y_k)\mathcal{P}(x_1, x_2,.......,x_{\mathcal{F}} | y_k)}{\sum_{j=1}^{\mathcal{M}} \mathcal{P}(y_j)\mathcal{P}(x_1, x_2,.......,x_{\mathcal{F}} | y_j)}
\end{equation}
where $\textbf{x}$ is $\mathcal{F}$ dimensional input feature vector of a sample input and $x_i (i\in \mathcal{F})$ is the $i^{th}$ number of feature of $\textbf{x}$. $y_k (k\in\mathcal{M})$ is the $k^th$ number of class. The assumption of Naive Bayes classifier is the conditional independent input features $x_i \in \mathcal{F}$  with class/output $y_k$, so that $\mathcal{P}(y_k | x_1, x_2,.......,x_{\mathcal{F}})$ can be represented as
\begin{equation}
	\mathcal{P}(y_k | x_1, x_2,.......,x_{\mathcal{F}})=\frac{\mathcal{P}(y_k)\prod_{i=1}^{\mathcal{F}}\mathcal{P}(x_i | y_k)}{\sum_{j=1}^{\mathcal{M}}\mathcal{P}(y_j)\prod_{i=1}^{\mathcal{F}}\mathcal{P}(x_i | y_j)}
\end{equation}
So, the classification rule for each new input $\hat{\textbf{x}}$ is
\begin{equation}
	\hat{y} = \arg\max_{k \in {1......\mathcal{F}}} \mathcal{P}(y_k)\prod_{i=1}^{\mathcal{M}}\mathcal{P}(\hat{x}_i | y_k)
\end{equation}
where $\hat{y}$ is the new classified output of the given input $\hat{\textbf{x}}$. In our case, the feature values are continuous, so the typical assumption is that the continuous values associated with each class are followed to Gaussian distribution.

\subsection{Multilayer Perceptron (MLP)}\label{appendix_MLP}
MLP is a discriminative classifier and makes a computational model based on the function and structure of biological neural networks. This classifier always tries to find out a pattern or complex relationship between inputs and outputs dataset. MLP consists of three layers: an input layer, a hidden, and an output. Each node in a layer is connected with every node of the following layer with a certain weight $w_{i,j}$. Now, our objective is to find the $w_{i,j}$. The function $f$ is represented in terms of weights. As finding the weights directly are often intractable, optimization technique is used for the same. The heart of this learning is based on the error, which is back propagated to find the optimum weights, in order to match the expected output. The degree of error of output node $j$ in the $n^{th}$ data sample, $e_j(n)$ can be represented as
\begin{equation}
	e_j(n)=t_j(n) - y_j(n)
\end{equation}
where $t$ and $y$ are the ground truth and the output value produced by perceptron.	The calculation of error minimization of entire output is given below.
\begin{equation}
	e_{min}(n)=\frac{1}{2}\sum e^{2}_j(n)
\end{equation}
The change in each weight is calculated using gradient descent, as represented in Equation \ref{change_weight}.
\begin{equation}\label{change_weight}
	\bigtriangleup w_{j,i}(n) = - \gamma \frac{\partial e_{min}(n)}{\partial v_j(n)}y_i(n)
\end{equation}
where $\gamma$ is the learning rate to select how quickly the weight gets converge and $y_i$ is the output of the previous node.

So, the classification rule for each new input $\hat{\textbf{x}}$ is
\begin{equation}
	\hat{y} = \arg\max_{k \in {1......\mathcal{M}}} y_k
\end{equation}
\bibliographystyle{IEEEtran}
\bibliography{mssi}

\end{document}